\documentclass[prl,twocolumn,floatfix,superscriptaddress,a4paper,nofootinbib]{revtex4}
\usepackage{amsmath,bm,graphicx}
\usepackage{amssymb}
\usepackage{amsfonts}
\usepackage{epstopdf}
\usepackage{mathrsfs}
\usepackage{xcolor}
\usepackage{latexsym}
\usepackage{hyperref}
\usepackage{subfigure}
\usepackage{bm}
\usepackage{natbib}
\usepackage{babel}
\usepackage{gensymb}

\usepackage{siunitx}

\newcommand*{\beginsupplement}{%
        \counterwithout{equation}{section}
        \setcounter{section}{0}
        \setcounter{table}{0}
        \renewcommand{\thetable}{S\arabic{table}}%
        \setcounter{figure}{0}
        \renewcommand{\thefigure}{S\arabic{figure}}%
        \setcounter{equation}{0}
        \renewcommand{\theequation}{S\arabic{equation}}%
        \setcounter{page}{1}
        \renewcommand{\thepage}{S\arabic{page}}%
     }

\begin{document}

\title{Zigzagging under extreme confinement}
\title{Zigzagging Diffusion and Non-Standard Transport in Particle-laden Nanopores Under Extreme Confinement}

\author{Andreas Baer}
\affiliation{Friedrich-Alexander-Universität Erlangen-Nürnberg, Department of Physics, Erlangen, Germany}
%
\author{Paolo Malgaretti}
\affiliation{Helmholtz Institute Erlangen-N\"urnberg for Renewable Energy (IET-2), Forschungszentrum J\"ulich, Germany}
\email[Corresponding author: ]{p.malgaretti@fz-juelich.de }
\author{Kevin H\"ollring}
\affiliation{Friedrich-Alexander-Universität Erlangen-Nürnberg, Department of Physics, Erlangen, Germany}
\author{Jens Harting}
\affiliation{Helmholtz Institute Erlangen-N\"urnberg for Renewable Energy (IET-2), Forschungszentrum J\"ulich, Germany}
\affiliation{Friedrich-Alexander-Universität Erlangen-Nürnberg, Department of Chemical and Biological Engineering, Germany}
\affiliation{Friedrich-Alexander-Universität Erlangen-Nürnberg, Department of Physics, Erlangen, Germany}
%
\author{Ana-Sun\v{c}ana Smith}
\affiliation{Friedrich-Alexander-Universität Erlangen-Nürnberg, Department of Physics, Erlangen, Germany}
\affiliation{Ru\dj er Bo\v skovi\' c Institute, Department of Physical Chemistry, Zagreb,Croatia}
\email[Corresponding author: ]{smith@physik.fau.de,\\asmith@irb.hr}

\begin{abstract}

Understanding transport subject to molecular-scale confinement is key to advancing nanofluidics, yet classical hydrodynamic laws often fail at these scales. Here, we study a model system: transport of toluene as a solvent and small fullerenes as model particles confined within alumina slit nanopores using molecular dynamics simulations. We find that toluene organizes into discrete layers whose commensurability with the pore width leads to a striking, non-monotonic, zig-zag dependence of transport coefficients on confinement. This layering drives oscillations not only in solvent diffusivity but also in flow velocity and permeability under pressure-driven conditions, breaking the expected scaling relations between diffusion, viscosity, and flow. Surprisingly, introducing a nanoparticle does not wash out these effects - although the fullerene perturbs local layering, the nanoparticle diffusivity retains a zig-zag dependence on pore width. Our results demonstrate how structural commensurability and interfacial effects dominate transport in nanoconfined liquids, and lead to important deviations from continuum expectations. These findings establish a microscopic basis for size-dependent transport in nanopores and highlight the need for beyond-hydrodynamic models in confined soft matter systems.
\end{abstract}
\maketitle

When liquids are confined within pores just a few molecular diameters wide, conventional assumptions about scale separation no longer hold.~\cite{achs_nalefd20_6937} Under such extreme confinement, transport becomes exquisitely sensitive to the structural details of the solute, solvent, and the confining walls~\cite{Charlaix2010,huber_soft_2015,Netz2021,Bocquet2023}.
This effect is especially striking in complex liquids like liquid crystals, which adopt distinct organizational patterns when confined by solid~\cite{zhang_columnar_2015,wiese_microfluidic_2016,wittmann_particle-resolved_2021} and liquid interfaces~\cite{peng_liquid_2018,chen_liquid_2021}. 
A similar behavior has also been reported for larger molecules, such as hydrocarbons~\cite{sivebaek_effective_2012},  organosilicons~\cite{matsubara_mechanism_2012}, and ionic liquids~\cite{Hollring2024b}. Lately, the effective diffusion coefficient of small molecules, such as toluene ~\cite{Baer2022}, and ions~\cite{Baer2022} has been reported to be sensitive to the degree of confinement. 

Recent results have also shown that the properties of ``simple liquids'', hard-sphere particles or even molecules such as water, undergo noticeable changes in the extreme confinement regime ~\cite{Franosh2022}. For example, a sharp transition from viscous to elastic response has been observed~\cite{khan_dynamic_2010}. Moreover, not only static properties but also dynamic properties exhibit changes such as non-linear rheology~\cite{bureau_nonlinear_2010}, spatially-dependent diffusivity~\cite{Hollring2023a}, and confinement-dependent viscosity~\cite{khan_viscosity_2014}.
This implies that, unlike in the meso- to macroscopic regime —- where the dynamic properties of liquids confined within pores typically vary monotonically with pore size -— in the highly confined regime, this relationship may change. Specifically, non-trivial dependence on pore size could emerge for structural and dynamic properties of liquid filling the pore. To what extent this affects the transport of solutes through these pores is yet to be understood.   

To address these questions, we perform an in-depth simulation study of the diffusion of fullerenes in toluene confined within a fully hydroxylated alumina slit nanopore. This model is chosen due to the non-polar nature of the interface, the solvent, and the solute, ensuring the dominance of weak dispersion interactions in the system. We observe the formation of multiple toluene layers at the alumina interface, whose commensurability with the confining geometry gives rise to a non-monotonic, zigzag dependence of the solvent diffusion coefficient $D$ as the pore width decreases. We, furthermore see that diffusion coefficients parallel and perpendicular to the pore wall differ significantly, and are altered from that of the bulk liquid. The associated effective viscosity obtained from equilibrium simulations differs from that inferred under steady-state flow 
Consequently zig-zag dependence of the effective P\'eclet number $Pe$ and the permeability on the pore width is observed, contrary to predictions of mesoscopic physics~\cite{Ghanbarian2022}.  Specifically, while the flow velocity $\bar{v}$ indeed exhibits non-monotonic trends, full compensation by the diffusion coefficient is not achieved. This complex behavior of the solvent ultimately reflects in the transport of particles in the fluid, in a manner that cannot be captured by expected continuum theories.

\begin{figure}
    \centering
    \includegraphics[width=0.45\textwidth]{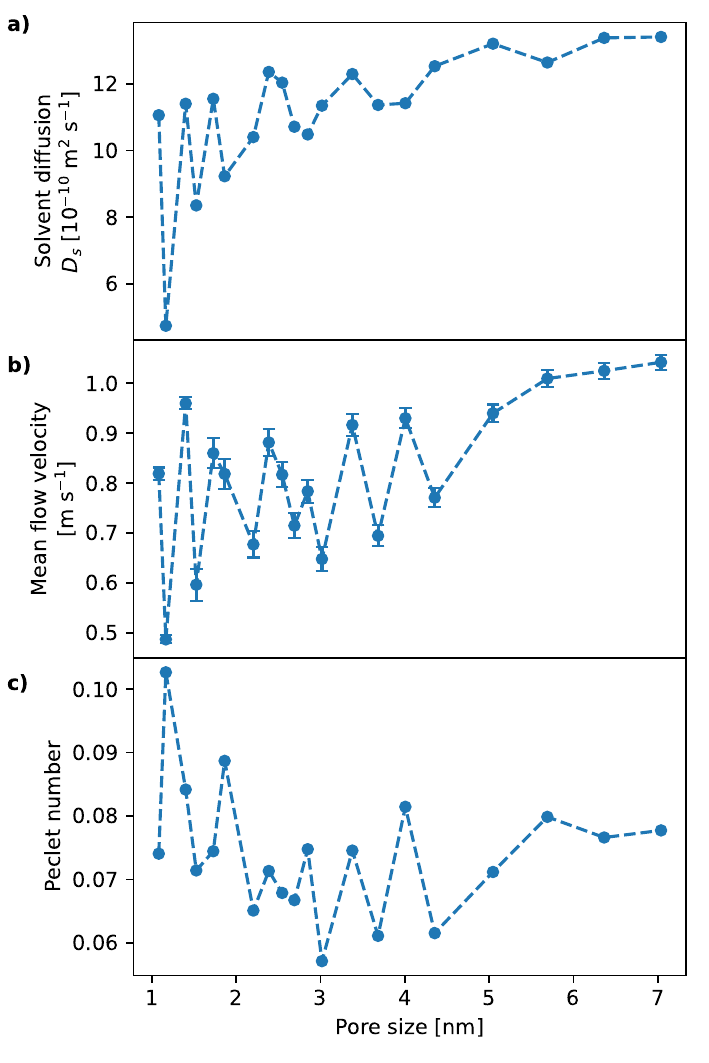}
    \caption{\textbf{Transport properties of toluene for different pore sizes.}
    \textbf{a)} The molecular self-diffusion coefficient $D_S$ of toluene in equilibrium shows a strong zig-zag dependence on the pore size for pore sizes $\lesssim \SI{2}{\nm}$. For larger pores the behavior is similar however less pronounced.
    \textbf{b)} When actively pulling the toluene through the pore at a constant force of \SI{20}{\kJ\per\mol\per\nm}, we observe a similar zig-zag behavior of the mean flow velocity $\bar v$, however it is present for pore sizes $\lesssim \SI{5}{\nm}$.
    \textbf{c)} The Peclet number $\textup{Pe} = \bar vd/D_S$ with the molecular size $d$ (taken as the diameter of the aromatic ring $d = \SI{0.66}{\nm}$) shows that the zig-zag behavior is not of similar magnitude for diffusion coefficient and mean velocity.
    }
    \label{fig:tol-diff}
\end{figure}

These findings stem from detailed MD simulations of toluene confined within a slit nanopore using the GROMACS simulation software~\cite{Gromacs1,Gromacs2,Gromacs3,Gromacs4,Gromacs5,Gromacs6,Gromacs7}.
For toluene, the OPLS-AA~\cite{OPLSAA1} force field is used
whereas for the C60 and C70, the atoms of the aromatic carbon of the OPLS-AA force field are used together with a structure from NMR data~\cite{Yannoni1991, C70source} following Ref.~\cite{Monticelli2012}.
The pore wall consists of hydroxylated bulk alumina ($\text{Al}_2\text{O}_3$) with the force field taken from~\cite{Vucemilovic2019, Vucemilovic2020}.
The system was built and equilibrated as in our previous work~\cite{Baer2022} (see Supplementary Information (SI) Section S1 for methodological details), whereby an alumina crystal was placed in an empty box to create a slit pore using periodic boundary conditions applied in all three dimensions. The void was then filled with toluene and, if applicable, a fullerene was added to the toluene. After energy minimisation, an NPT equilibration was performed with semi-isotropic coupling, where only the pore width was adjusted~\cite{Huang2008, Chinappi2010} to attain the correct density at ambient pressure. Finally, an NVT equilibration step was conducted under production conditions.
In all simulations, the temperature was controlled by the BDP velocity rescaling thermostat \cite{Bussi2007} with a coupling time of 1.0~ps, to keep the system at 293.15~K, while in the NPT run, the C-rescale barostat was used to control the pressure with a coupling time of 5.0~ps.

We first calculate the dependence of the diffusion coefficient parallel to the pore surface of the toluene, averaged over all molecules, as a function of the pore width (Fig.~\ref{fig:tol-diff}{a}; methodology in SI Section S2). For pore widths $w\simeq 5$~$\text{nm}$ i.e., in the extreme confinement scenario, the diffusion coefficient has a zig-zag dependence superimposed on an overall decrease as the pore becomes more and more narrow. This is consistent with the behavior observed in Lennard-Jones fluids~\cite{beer_non-monotonic_2012}. As $w\gg 3$~$\text{nm}$, the amplitude of oscillations decays until it vanishes for even larger pore sizes. 
The peak-to-peak distance of the zig-zag pattern is roughly 0.3~$\text{nm}$, which is consistent with the thickness of a toluene molecule along its smallest dimension. 

These results are consistent with the idea that in commensurate packing, in which confinement length closely matches an integer multiple of the solvent diameter, molecules tend to arrange into well-defined layers while deviations from this result in a more “frustrated” and interlocked liquid structure.~\cite{Lowen1996, Franosh2022}. It is somewhat surprising that, under constant pressure within the pore, and with the still preserved liquid state, changes in the pore width that are much smaller than the molecular size affect transport so strongly.  

\begin{figure}[!h]
    \centering
    \includegraphics[width=0.8\textwidth]{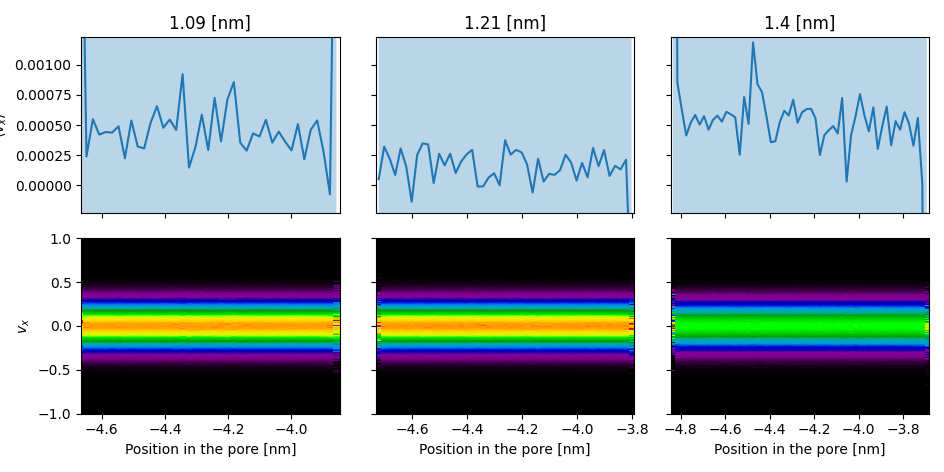}
    \caption{\textbf{Fluid velocity induced by pulling forces}. Constant velocity profiles accross the pore when toluene is pulled actively with a force constant of \SI{10}{\kJ\per\mol\per\nm}.
    \textbf{Top:} mean velocity and
    \textbf{bottom:} full velocity distribution as a function of the position in the pore.
    The mean velocity for the three pore sizes are \SI{0.44}{\m\per\s} (\SI{1.09}{\nm}), \SI{0.17}{\m\per\s} (\SI{1.21}{\nm}), and \SI{0.53}{\m\per\s} (\SI{1.40}{\nm}).
    This resembles the same trend as the diffusion coefficient of toluene in the three smallest pores: \SI{7.4e-10}{\m\squared\per\s} (\SI{1.09}{\nm}),  \SI{5.6e-10}{\m\squared\per\s} (\SI{1.21}{\nm}), and  \SI{7.7e-10}{\m\squared\per\s} (\SI{1.40}{\nm}).
    }
    \label{fig:pull-velocity}
\end{figure}

\begin{figure}[!h]
    \centering
    \includegraphics[width=0.65\textwidth]{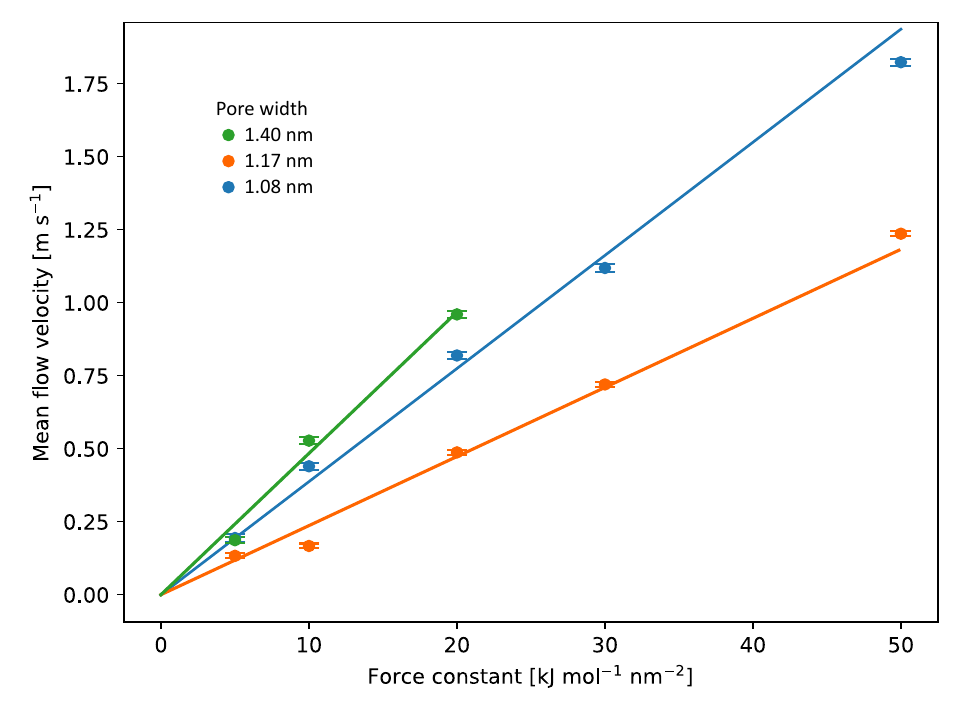}
    \caption{\textbf{Linear response}. Mean flow velocity as a function of the applied force constant for different pore sizes. Error bar are standard deviations, from the average performed over the velocities of all toluene molecules in the pore, over all time-frames of the production run.}
    \label{fig:mean-velocity}
\end{figure}

To assess whether this result is compatible with mesoscopic physics ~\cite{Chantal2024, Speck2025}, we examine the relationship between diffusivity and flow velocity, captured by the  P\'eclet number $\text{Pe}$ ($Pe = \bar{v} d/D$) at fixed pressure gradient, using the diameter of the aromatic ring of toluene as the characteristic length $d = 0.66\ \text{nm}$. For that purpose, we performed a series of non-equilibrium molecular dynamics simulations to capture the effect of applying external forces of increasing magnitudes to toluene (Fig.~\ref{fig:pull-velocity}), after confirming linear response in bulk liquids for the same force range (Fig.~\ref{fig:mean-velocity}). Consistent with observations of non-monotonous diffusivity, the zig-zag dependence is observed for the net flow of toluene (Fig.~\ref{fig:tol-diff}b). However, in the extreme confinement, the diffusivity and flow velocity oscillations are not compensatory, so that the P\'eclet number also zig-zags, which is a clear deviation from the mesoscopic prediction (Fig.~\ref{fig:tol-diff}c). 

\begin{figure*}
    \centering
    \includegraphics[width=\textwidth]{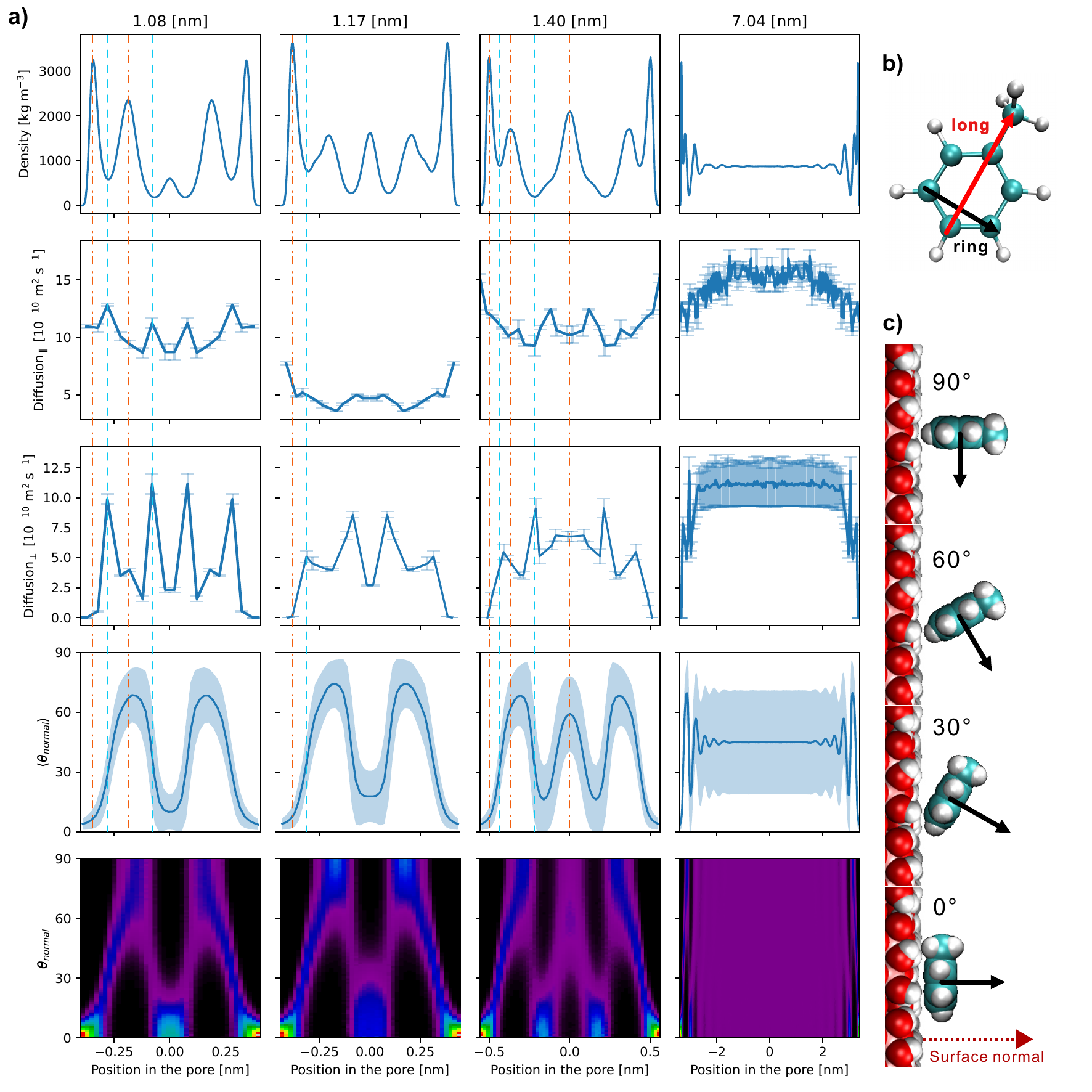}
    \caption{\textbf{State of toluene in the pore.}
    \textbf{a)} For four different pore sizes (columns, headers give the approximate width of the pore void), several quantities are presented as functions of the position in the pore (vertical lines as a visual guide for the alignment of main features):
    \textbf{Density:} toluene molecular COM density distribution,
        \textbf{Diffusion:} toluene diffusion coefficient in parallel ($\parallel$) or perpendicular ($\perp$) direction to the pore wall
    \textbf{Mean orientation:} mean (solid line) and standard deviation (shaded area), of the angle, the ``normal'' vector encloses with the surface normal.
    \textbf{Orientation distribution:} histogram of the toluene orientation as the angle the ``normal'' vector encloses with the surface normal.
    \textbf{b)} A toluene molecule can be regarded a platelet with approximate sizes of $0.83\text{nm}\times 0.66\text{nm}\times 0.33\text{nm}$ and mass $92.141\text{u}$. It has two native vectors describing the molecular orientation: the vector ``long'' connects the carbon atom outside the aromatic ring with the carbon atom farthest away, being the longest internal axis of the molecule. The vector ``ring'' connects two atoms in the aromatic ring and is (in the absence of internal vibrations) orthogonal to the vector ``long''.
    \textbf{c)} Different alignment configurations of a toluene molecule near a surface are depicted in a side view on both the toluene molecule and the wall.
    An additional orientation vector ``normal'' is sketched.
    The vector is normal to the plane of the aromatic ring and thus orthogonal to both vectors ``ring'' and ``long''.
    The normal of the wall surface is also depicted (dashed red arrow), being the reference for angle calculations.
    All angles in this manuscript are calculated as the angles that a specific vector encloses with the surface normal vector.
    The angles given in the sketches refer to the ``normal'' vector. A perpendicular alignment of toluene appears for $\theta_{long} = \theta_{ring} = 90 \degree$ while the parallel orientation is observed for $\theta_{normal} = 0 \degree$. Due to the symmetry of the molecule, angles between $\theta_{normal} = 90 \degree$ to $180 \degree$ can be mapped back to the range $90 \degree$ to $180 \degree$.}
    \label{fig:orientation}
\end{figure*}

To relate this behavior to effects occurring on the molecular scale, we thoroughly analyzed the equilibrium local structuring of toluene within the pore (top row of Fig.~\ref{fig:orientation}a and Fig.~\ref{fig:solute-dens}). In the extreme confinement regime (first three columns of Fig.~\ref{fig:orientation}a), several layers of toluene are clearly discerned within the pore while no ``bulk'' solvent region, characterized by a homogeneous toluene density profile, is observed. For weaker confinement (rightmost column in Fig.~\ref{fig:orientation}a), a bulk region in the middle of the pore develops beyond four interfacial layers.

We furthermore calculate the spatially resolved diffusion coefficients of toluene (second and third row of Fig.~\ref{fig:orientation}a). Remarkably, the density and diffusivity profiles seem to be weakly correlated: strong density oscillations are inversely related to positions but not amplitudes of the peaks in diffusivity, while the baseline of parallel diffusivity is strongly shifting (as captured by its average (Fig.~\ref{fig:tol-diff})). 

The oscillations in density are more intense in the profile of particle transport in the direction perpendicular to the pore compared to parallel diffusion profiles with local maxima in density corresponding to local lows in particle mobility. 
The precise relation is however complex, especially in the regions between the extrema of density and no simple relationship between the absolute density and the perpendicular mobility can be derived \cite{Hollring2023a}. While in parallel direction, diffusivity attens the bulk value at bulk density, the perpendicular diffusion coefficient remains sensitive to the confinement even at interemediate pore widths. 
Therefore, our data shows not only that it is not possible to extract the transport coefficient from the density profile, but also that it is not trivial at all to make even an ``educated guess'' about the local diffusivity by inspecting the density profile. 

\begin{figure*}[!h]
    \centering
     \includegraphics[width=0.2\textwidth]{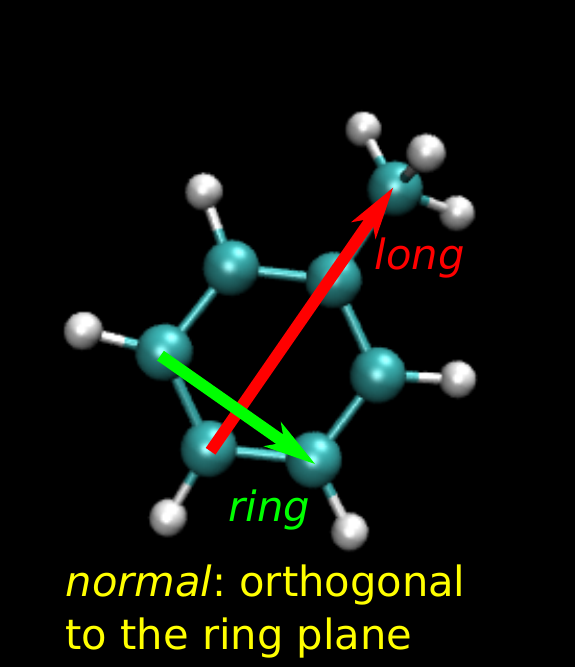}
    \includegraphics[width=0.9\textwidth]{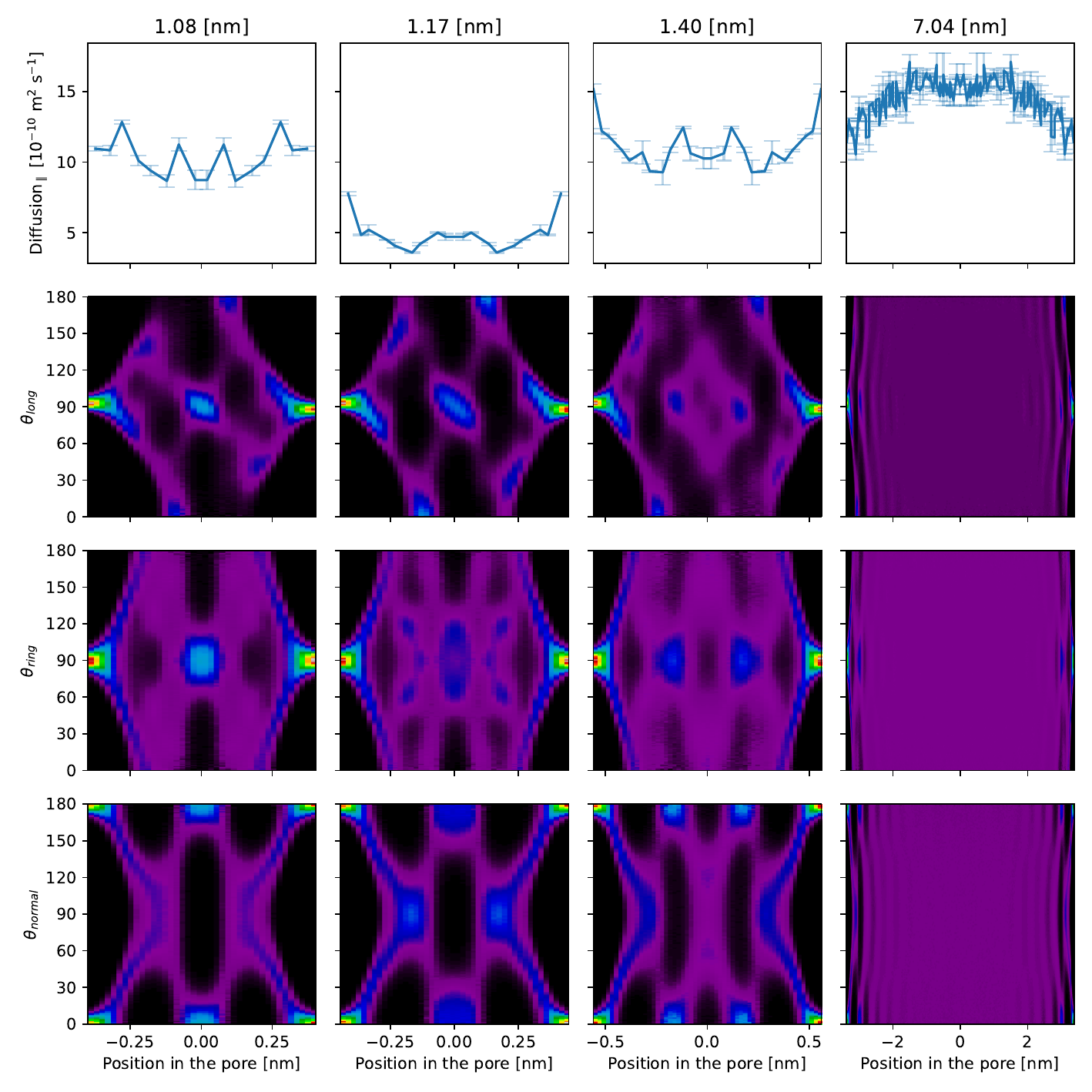}
    \caption{\textbf{Distribution of toluene. orientational degrees of freedom.} Toluene molecule with sketches of the different orientation vectors.
    The normal vector is the normal to the plane of the aromatic carbon ring. Bottom: Full orientation profiles of all three orientation vectors.}
    \label{fig:orientation-full}
\end{figure*}

Interestingly, a correlation of diffusivity was found with the average orientation of the toluene in the pore (fourth and fifth row in Fig.~\ref{fig:orientation}a) and the distribution of the three toluene orientation angles (Fig.~\ref{fig:orientation-full}). The orientation relates to the effective hydrodynamic radius of toluene molecules for transport in a parallel or perpendicular direction, but it is also driven by confinement. For example, toluene molecules lie flat at the wall surface due to the hydrophobicity of the hydroxylated alumina and the non-polar nature of the toluene, while for pores with $d\approx 3~\text{nm}$, toluene explores all orientations as its local environment is that of the bulk liquid.

\begin{figure}
    \centering
    \includegraphics[width=0.45\textwidth]{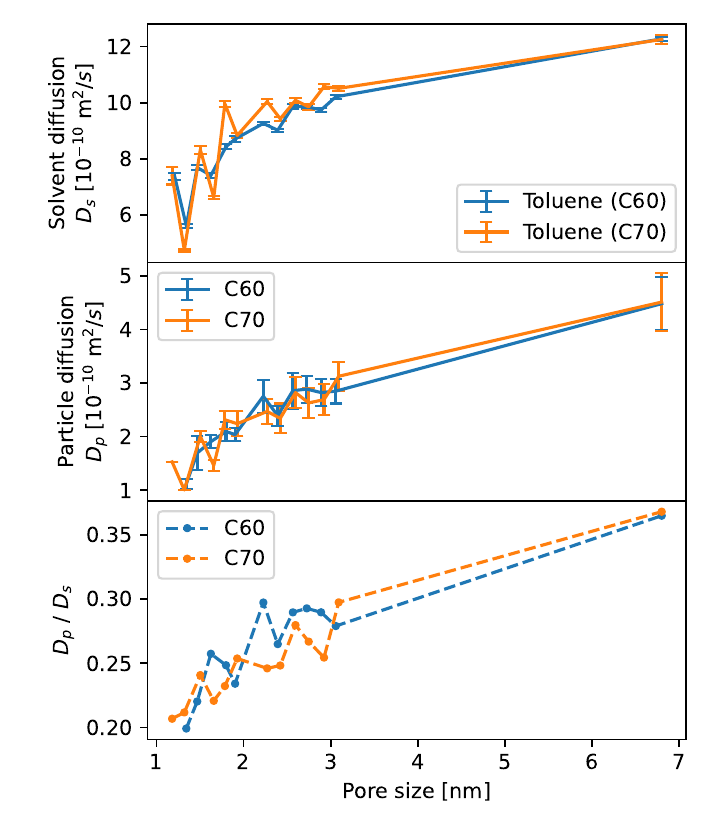}
    \caption{Pore size dependence of transport through the pore.
    Shown is the mean lateral diffusion of the molecular centers of mass for
    \textbf{a)} the solvent for the two different systems. The zig-zag for small pores is apparent.
    \textbf{b)} The solute.
    \textbf{c)} The ratio of solute to solvent is depicted to correct the solute diffusion for a change in solvent viscosity.
    All lines are guides to the eye.
    }
    \label{fig:diff-vs-size}
\end{figure}

In the next step, we study how the non-monotonic, zig-zag dependence of the solvent transport coefficients affects the transport of solutes. For this, we add C60 and C70 fullerenes into the pore and show that the zigzagginess of toluene diffusivity with the pore size persists even in the presence of dissolved particles (cf.~Fig.~\ref{fig:tol-diff}a and top panel in Fig.~\ref{fig:diff-vs-size}), although the solvent layering is somewhat disrupted (Fig.~\ref{fig:solute-pmf}). These results demonstrate that under strong confinement, classical relationships between diffusion, viscosity, and permeability no longer hold quantitatively and that such deviations persist even in particle-laden fluids.

Strikingly, the fullerenes themselves exhibit a comparable zig-zag dependence (middle panel in Fig.~\ref{fig:diff-vs-size}). Although C60 (radius $\approx 0.52$~nm) and C70 (ellipsoidal, long axis $\approx 0.61$~nm) are fully accommodated even in most narrow pores, their diffusivity increases fivefold between pore widths of 1 and 7~nm, compared to a twofold increase for toluene. This is likely due to the complex interplay of the reduced surface transport of fullerenes~\cite{Baer2022} with the strong confinement of the liquid, causing a stronger effective friction (SI Section S6). The difference between C60 and C70 diffusivity demonstrates the large sensitivity of transport properties to solute properties. It stems from the slightly elliptical shape of C70, which can adjust its orientation to the local solvent environment.

Moreover, while for toluene the majority of the change in the diffusivity occurs for pores of 1-3~nm, the fullerenes remain sensitive to confinement across the entire studied range of pore widths. Furthermore, under strong confinement, the solute-to-solvent diffusivity ratio varies non-monotonically (bottom panel of Fig.~\ref{fig:diff-vs-size}). This behavior contrasts sharply both bulk conditions -- where fullerene diffusivity scales predictably with toluene viscosity and particle size via the Stokes–Einstein relation~\cite{Baer2024} -- and weaker confinement, where the emergence of a bulk-like region at the pore center restores classical transport laws. This indicates that classical hydrodynamic relations not only fail but cannot be recovered via an effective radius. Transport in this regime must be treated on fundamentally different grounds.

In summary, we have investigated the transport properties of toluene and suspended C60 and C70 fullerenes under extreme confinement, using this as a model system where the discrete molecular structure of the fluid becomes important. In this regime, the diffusivity of toluene exhibits a pronounced, non-monotonic zig-zag dependence on pore size, driven by commensurability effects between solvent layering and pore geometry. This behavior disappears when the pore size exceeds roughly ten molecular layers, highlighting its confinement-specific origin.

Remarkably, the zigzagging trend persists under non-equilibrium conditions, where flow is driven by an external pressure gradient. In this case, the standard proportionality between diffusivity and flow velocity breaks down, leading to a similarly non-monotonic dependence of both permeability and the effective Péclet number on pore size. This behavior remains robust in the presence of suspended nanoparticles:  C60 and C70 locally disrupt solvent layering which is reflected in subtle modifications of the solvent's characteristic zig-zag diffusivity dependence on the pore width. However, their own diffusivities also display a zigzag dependence on confinement. 

We conclude that the zigzag dependencies observed systematically arise from a complex interplay between solvent transport, molecular layering near the pore walls. The layering manifests as density oscillations whose commensurability with the pore width varies, thereby influencing the overall molecular mobility throughout the pore. These density profiles reflect the underlying solvent-wall effective potential, which in turn induces drifts within the solvent, the consequence of which is a complex coupling of diffusivity and density profiles. Transport behavior is further modulated by a spatially varying effective hydrodynamic radius of the solvent, stemming from orientational preferences of the solvent molecules relative to the axis of motion. Such a complex solvent environment persists even in the presence of solutes, underscoring the need to go beyond classical transport relations when modeling transport under strong confinement.

\textbf{Acknowledgements:} 
We thank the Deutsche Forschungsgemeinschaft (DFG, German Research Foundation) project 416229255 SFB 1411 Design of Particulate Products for financial support. The computations for this project were performed at the Friedrich-Alexander-Universität Erlangen-Nürnberg (FAU) center for Advanced Computing RRZE and with the support of NHR@FAU under the project b171dc. Further resources were provided by the University of Zagreb Computing Centre - SRCE. Input files along with raw data are provided as a ZENODO archive with the DOI: \hyperlink{https://doi.org/10.5281/zenodo.15535557}{10.5281/zenodo.15535557}\cite{baer_2025_15535558}.

\bibliography{bibliography}

\beginsupplement

\widetext
\begin{center}
\newpage
\textbf{\large Supplementary Information \\ 
\vspace{5mm}
Zigzagging Diffusion and Non-Standard Transport in Particle-laden Nanopores Under Extreme Confinement}\\
\vspace{5mm}

\medskip
Andreas Baer$^{1}$, Paolo Malgaretti$^{2,*}$, Kevin H\"ollring$^{1}$, Jens Harting$^{2,3,1}$ and Ana-Sun\v{c}ana Smith$^{1,3,4*}$

\medskip

\textit{\small{$^{1}$Friedrich-Alexander-Universität Erlangen-Nürnberg, Department of Physics, Erlangen, Germany}}\\
\textit{\small{$^{2}$Helmholtz Institute Erlangen-N\"urnberg for Renewable Energy (IET-2), \\Forschungszentrum J\"ulich, Erlangen, Germany}} \\
\textit{\small{$^{3}$}Friedrich-Alexander-Universität Erlangen-Nürnberg, \\Department of Chemical and Biological Engineering, Erlangen, Germany}\\
\textit{\small{$^{4}$} Ru\dj er Bo\v skovi\' c Institute, Department of Physical Chemistry, Zagreb,Croatia}\\ 
\vspace{2mm}
\texttt{\small{$^{*}$p.malgaretti@fz-juelich.de}}
\texttt{\small{$^{*}$smith@physik.fau.de, asmith@irb.hr}} 
\end{center}
\endwidetext
\subsection{S1. Simulation methods}

Following a similar procedure as in our previous work~\cite{Baer2022}, all the simulations are performed with the 2021.3 version of Gromacs~\cite{Gromacs1,Gromacs2,Gromacs3,Gromacs4,Gromacs5,Gromacs6,Gromacs7} with the leap-frog integrator and a time step of \SI{1}{\fs}.
All bonds involving hydrogen atoms are
constrained using the LINCS algorithm~\cite{LINCS}.
Short-range non-bonded interactions are treated with a Lennard-Jones potential  switched smoothly to zero between \num{0.9} and \SI{1.2}{\nm} \cite{Christen2005}, while long range electrostatic interactions are calculated with the PME technique \cite{PME1,PME2}.
In all simulations, the temperature and, if applicable, pressure coupling occurs every ten steps, which is also the frequency at which the center-of-mass motion of the whole system is removed.

The NPT equilibration is conducted for \SI{5}{\ns} at ambient pressure and \SI{293.15}{\K}.
The pressure coupling is applied only in the direction orthogonal to the pore wall, such that only the width of the pore is adjusted~\cite{Huang2008, Chinappi2010}.
The velocities are initialized according to a Maxwell distribution at the desired temperature of \SI{293.15}{\K}.
After NPT equilibration, for pores with a width of at least $\approx \SI{3}{\nm}$), we verify that the density of toluene in the center of the pore attains values within \SI{1}{\percent} of the experimental bulk density of toluene of \SI{867.1}{\kg\per\meter\cubed} \cite{Daridon2018}. This coincides with the estimated statistical uncertainty of density in unbounded toluene simulations in same thermodynamic conditions. For pores where no bulk liquid is present, the deviations from the bulk value increase as the pore width decreases. For example, in pores \SI{2}{\nm} wide, the density deviates less than $5 \%$ from the bulk value, while even for the smallest pores the deviation is only about $15 \%$.

The positions of all atoms in the system are recorded every \si{\ps} and the diffusion coefficient, density and orientation are all deduced from this data. In those simulations, where the toluene is actively pulled through the pore, a constant force is applied to all toluene molecules simultaneously, mimicking a constant pressure drop across the pore.
The velocity of the whole system is reported every \SI{2}{\ps}.
In order to obtain the toluene flow velocity, the center-of-mass motion of the pore wall (the hydroxylated alumina) is calculated along with the velocity of the toluene molecules relative to the pore wall.

\subsection{S2. Determining profiles of diffusion coefficients for transport parallel and perpendicular to pore walls}

Due to the lack of symmetry, we restrict our analysis here to the interface-parallel direction. 
Based on the time-dependent positions of particles $\vec{r}_i(t)$ and the coordinate in interface-parallel direction $\vec{r}_{i,\parallel}(t)$ , the interface-parallel mean squared displacement $\mathrm{MSD}_\parallel$ is calculated as the following average:
\begin{equation}
    \mathrm{MSD}_\parallel(\Delta t) = \langle(\vec{r}_{i,\parallel}(t+\Delta t)-\vec{r}_{i,\parallel}(t))^2\rangle_{i, t} \label{eq:parallel_MSD}
\end{equation}
From this, we can then extract the parallel diffusion coefficient $D_\parallel$ via a linear fit of the linear region of the $\mathrm{MSD}_\parallel(\Delta t)$ through the relationship:
\begin{equation}
   2 \times n \times D_\parallel = \lim_{\Delta t\to\infty} \frac{d}{d (\Delta t)} \mathrm{MSD}_\parallel(\Delta t).
\end{equation}
Here, $n$ denotes the number of dimensions considered, which in the case of a two-dimensional subspace of the interface-parallel position is $n=2$.

To obtain the spatially dependent diffusion profile $D_\parallel(z)$ with $z$ being the direction perpendicular to the pore walls, we use freely available, custom tools built on top of the GROMACS API, as in \cite{Hollring2023a,Hollring2024b}. Specifically, we split the pore into slices of thickness $L=\SI{0.1}{\nano\meter}$, starting at position $z_i$, such that the slice $i$ is represented by the subspace $z\in [z_i, z_i+L_i]$.
As these subspaces do not evenly divide the pore width, the last, incomplete such space is discarded. Further, the resulting profile is mirrored across the center of the pore to account for the intrinsic symmetry of the system.
The trajectories of particles are then split for each slice into sub-trajectories, where the particle was continuously within the slice. 
Whenever the center of mass of the particle leaves the slice and returns back into it, it is then treated as a new sub-trajectory. 
The calculation of $\mathrm{MSD}_{\parallel,i}$ is then performed as in Eq.~\ref{eq:parallel_MSD} but restricted to these sub-trajectories.
Hence, the average $\langle\cdot \rangle_{i,t}$ is calculated as the time average only considering events where the particle remains continuously within the slice $i$ from $t$ to $t+\Delta t$. 
Then, the mean diffusion coefficient within the slice $D_{\parallel,i}$ becomes
\begin{equation}
   2 \times n \times D_{\parallel,i} = \lim_{\Delta t\to\infty} \frac{d}{d (\Delta t)} \mathrm{MSD}_{\parallel,i}(\Delta t).
\end{equation}
In particular, we calculate the approximate slope limit via a linear fit in the occurring linear regime of $\mathrm{MSD}_{\parallel,i}(\Delta t)$. At longer times, the statistics are affected due to the confinement to one thin slice.


For the perpendicular diffusion, we use the so-called Simple Particle Model (SPM) for extracting diffusion coefficients from the distribution of life times of a particle within each slice. \cite{Hollring2023a} As the pore effects are strong, we resort to the version of the model that accounts for the impact of drift resulting from the effective background potential, following the protocol fully validated in our previous publication. Specifically, the distribution of life times was fitted by its analytic prediction. When necessary due to the finiteness of the trajectories, the distributions were extended by an exponential curve to reach to the long-time limits.

The perpendicular diffusion coefficient was then respectively calculated via the relationship
\begin{equation}
   D_\perp = \frac{1}{12} \times \frac{L^2}{\langle\tau\rangle}.
\end{equation}
The correction for drift (SPM+d approach in \cite{Hollring2023a} for details) was then applied to our resulting diffusivity to obtain reliable local perpendicular diffusion coefficients.
We used the corrected result from the fit of the analytic model as a basis for our plots. Standard deviation was calculated from diffusivities calculated without drift correction, without long-time limit extension of the distribution and the fully corrected result. 

\subsection{S3. Density profiles of fullerenes in narrow pores}
\begin{figure}[!h]
    \centering
    \includegraphics[width=0.65\textwidth]{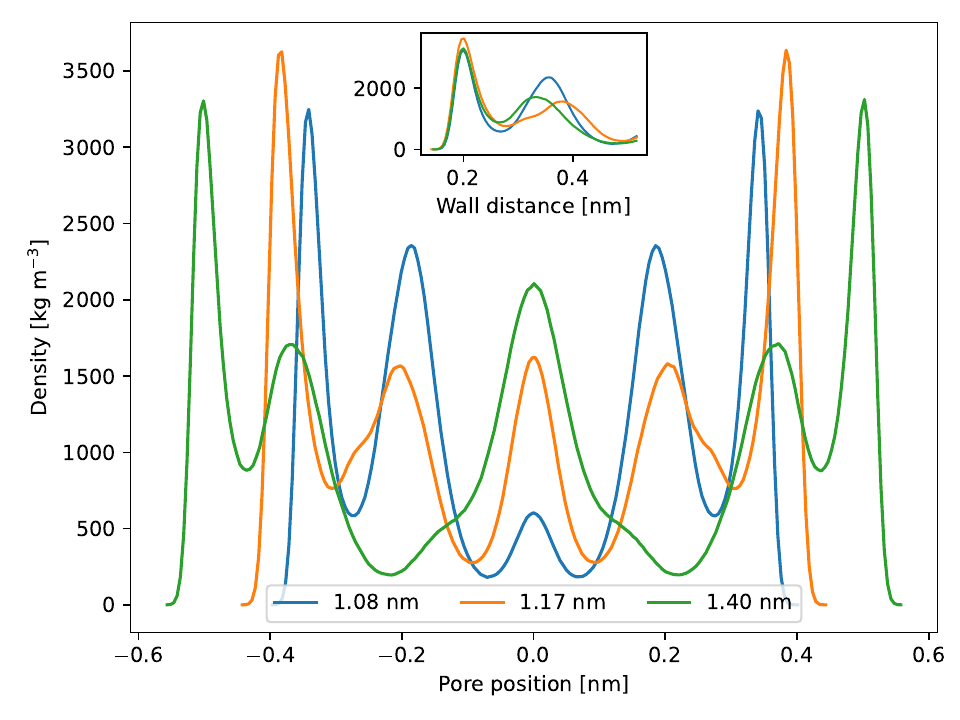}
    \caption{\textbf{Solvent density in the three smallest pores as a function of the distance from the center of the pore.} 
    Strong differences in the layering are apparent.
    While the first peak is indistinguishable (see inset), the second one is either pronounced or smeared out, depending on the most occupied toluene configurations in the pore.}
    \label{fig:solute-dens}
\end{figure}

\begin{figure}[!h]
    \centering
    \includegraphics[width=0.45\textwidth]{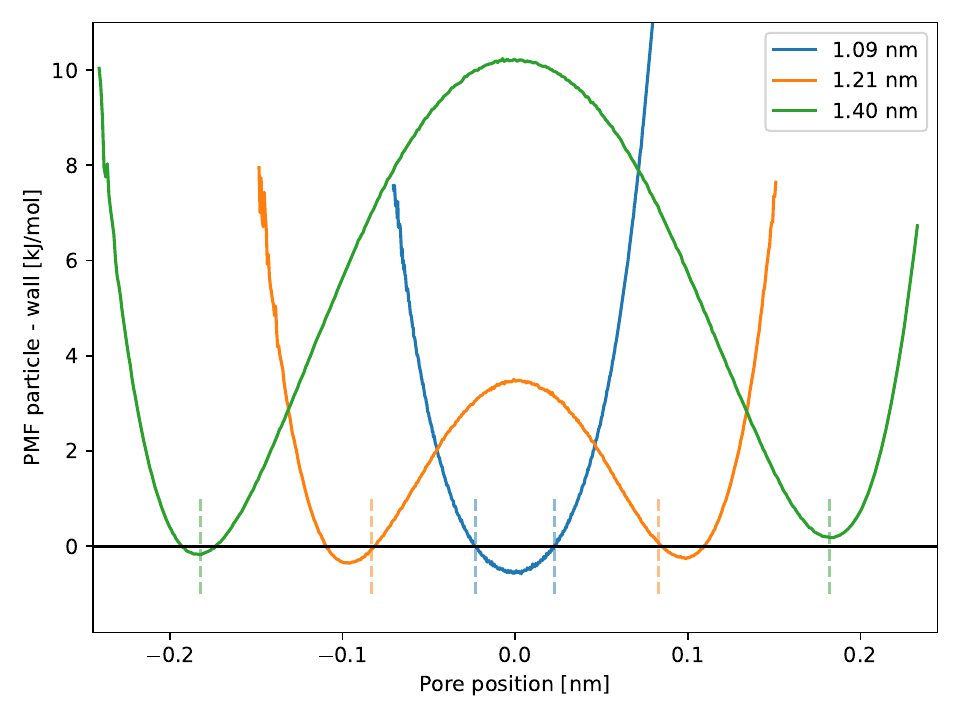}
    \includegraphics[width=0.45\textwidth]{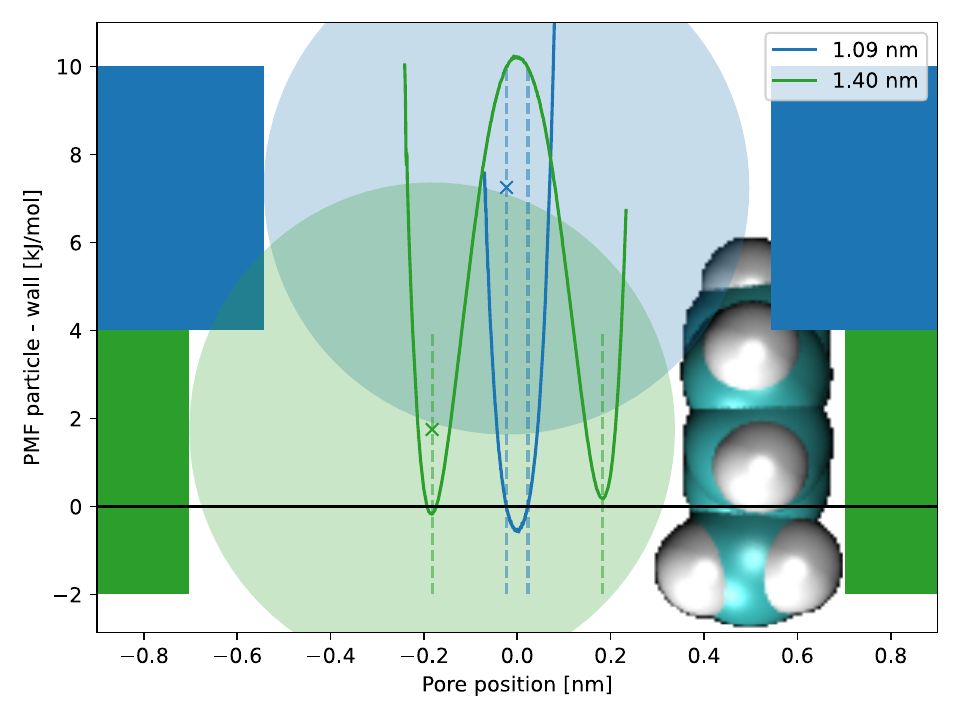}
    \caption{\textbf{Fullerene in the toluene-filled pore}. Potential of mean force for a C60 in the three smallest pores.
    The dashed lines give the positions, where the particle is in contact with the wall.
    \textbf{Left:} Potential of Mean Force for the positioning of the center of mass (COM) of the fullerene in pores of different widths.
    \textbf{Right:} Sketch of a pore of a width 1.09 nm (blue) and 1.4 nm (green). Light blue and green circles represent a C60 fullerene in such pores, drawn to scale. In the narrow, blue pore, fullerene expels all solvent molecules, while in the green pore, there is just enough space for 1 toluene molecule. This drives a transition from a single-well to a double-well potential of mean force for a fullerene COM in the pore.}
    \label{fig:solute-pmf}
\end{figure}

While the smallest pore only allows for a single minimum in the middle of the pore, the larger pores allow for two minima at the walls (see figs.~\ref{fig:solute-dens} and \ref{fig:solute-pmf}).
For the intermediate pore ($1.21\text{nm}$) parts of a toluene molecule can squeeze in the empty space between the fullerene and the other wall.
For the larger pore ($1.40\text{nm}$), a full layer of toluene fits in the empty space, giving rise to the strong potential barrier (about $25\text{kJ}\times \text{mol}$) between the two minima.

\widetext

\vspace{45mm}

\end{document}